\input phyzzx
%\input epsf.tex
%\epsfverbosetrue
\hfuzz 50pt

\def\cM {{\cal{M}}}

%%%%%%%
\font\mybb=msbm10 at 12pt
\def\bb#1{\hbox{\mybb#1}}

\def\bR {\bb{R}}

\def\bfx {{\bf{x}}}
\def\bfy {{\bf{y}}}

\def\bfomeg{\omega\kern-7.0pt\omega}
\def\bfOmeg{\Omega\kern-8.0pt\Omega}

%%%%%%%%%%%%

%%%%%%%%%%%%%%%%%%%%%%%%%%%%%%%%%%%%%%%%%%%%%%%%%%%%%%%%%
\REF\coles{ RA Coles and
 G Papadopoulos, {\it The Geometry
of the One-dimensional
Supersymmetric Non-linear Sigma Models},
Class. Quantum Grav. {\bf 7} (1990)
427-438.}

\REF\stelle{GW Gibbons,
G Papadopoulos  and KS Stelle,
{\it HKT and OKT Geometries on
Soliton Black Hole Moduli Spaces},
 Nucl.Phys. {\bf B508} (1997)623; hep-th/9706207.}
 
 \REF\stroma{J Michelson and A Strominger, {\sl Superconformal 
Multi-Black Hole
Quantum Mechanics},  JHEP 9909:005, (1999):  hep-th/9908044.}

\REF\gutpapa{J Gutowski and G Papadopoulos, {\it The dynamics of very 
special black holes}
 Phys.Lett. {\bf B472}:45-53, (2000): hep-th/9910022.} 
 
 \REF\stromb{A Maloney, M Spradlin and
 A Strominger, {\sl Superconformal
Multi-Black Hole Moduli Spaces 
in Four Dimensions}, hep-th/9911001.}

\REF\gutpapb{J Gutowski and G Papadopoulos, {\it Moduli spaces for 
four-dimensional and five-dimensional black
holes} Phys.Rev.{\bf D62}:064023, (2000): hep-th/0002242.} 

\REF\hr{  SJ Gates, Jr., CM Hull  and M Ro\v cek, {\it Twisted multiplets and 
new supersymmetric nonlinear sigma
models},  Nucl.Phys. {\bf B248}:157, (1984).}

\REF\howepapc{PS Howe and G Papadopoulos, {\it Ultraviolet 
behavior of two-dimensional 
supersymmetric
nonlinear sigma models}, Nucl.Phys. {\bf B289}: 264, (1987); 
{\it Further remarks on the geometry of two-dimensional nonlinear
sigma models},  Class.Quant.Grav. {\bf 5}:1647, (1988).}

\REF\yano{K Yano, {\it Differential Geometry on complex and almost complex spaces}, 
Pergamon Press, Oxford (1965).}

\REF\howepap{PS Howe and
G Papadopoulos, {\it Twistor Spaces
for HKT Manifolds}, Phys. Lett.
{\bf B379} (1996) 80; hep-th/9602108.}

\REF\poona{G Grantcharov
and Y-S Poon, {\it Geometry of Hyper-K\"ahler
Connections with Torsion}, Commun.Math.Phys. {\bf 213} (2000) 
19-37: math.dg/9908015. }

\REF\poonb{G Grantcharov, G Papadopoulos and Y-S Poon, {\it Reduction 
of HKT structures},
math.DG/0201159.  }

\REF\tvp{P Spindel, A Sevrin,
W Troost and A Van Proeyen,
{\it Extended Supersymmetric Sigma
Models on Group Manifolds. 1. The
Complex Structures}
Nucl. Phys. {\bf B308} (1988) 662. }

\REF\opap { A Opfermann and G Papadopoulos, 
{\it Homogeneous HKT and QKT
manifolds}: math-ph/9807026.}

\REF\dofi{ IG Dotti and A Fino, {\it Hyperk\"ahler torsion structures 
invariant by nilpotent Lie groups}: math.DG/0112166.}

 \REF\howepapb{PS Howe,
 A Opfermann and G Papadopoulos,
 {\it Twistor Spaces for QKT Manifolds},
 Commun.Math.Phys. {\bf 197}(1998) 713: hep-th/9710072.}

 \REF\ivanov{S Ivanov, {\it Geometry
 of Quaternionic K\"ahler
 connections with torsion}, to appear in Journal Geom. Phys.: math.DG/0003214.}
 
 \REF\ivanovd{S Ivanov and I Minchev, {\it 
 Quaternionic K\"ahler and hyperK\"ahler manifolds with torsion and twistor spaces},
 math.DG/0112157. }
 
 \REF\giba{GW Gibbons and PJ Ruback, {\it The motion of extreme
 Reissner Nordstr\"om black holes in the low velocity limit},
 Phys. Rev. Lett. {\bf 57} (1986) 1492.} 

\REF\gutpapc{J Gutowski and G Papadopoulos, {\it Three body interactions,
 angular momentum and black hole
moduli spaces}: hep-th/0107252. }

\REF\swann { Y-S Poon and A Swann  
{\it Superconformal symmetry and hyperKaehler manifolds with torsion}:
     math.DG/0111276. }

\REF\schwarz{J Schwarz, {\it Covariant field equations of chiral N=2  D=10
supergravity}, Nucl. Phys. {\bf B226} (1983) 269.}

\REF\howewest{ PS Howe and PC West, {\it The complete N=2 D=10 supergravity},
 Nucl. Phys.
{\bf B238} (1984) 181.}

\REF\west{
ICG Campbell and PC West, {\it N=2 D = 10 nonchiral supergravity and 
its spontaneous
compactification}, Nucl.Phys. {\bf B243} (1984) 112.} 

\REF\ivapapa{ S Ivanov and G  Papadopoulos, {\it A no-go theorem for
 string warped compactifications},  Phys.Lett. {\bf B497}:309-316, 
 (2001): hep-th/0008232.} 
 
\REF\ivapapb{
S Ivanov and G Papadopoulos,  {\it Vanishing theorems and 
string backgrounds},  Class.Quant.Grav. {\bf 18}:1089-1110, 
(2001): math.dg/0010038.} 

 \REF\ivanovb{B. Alexandrov and S. Ivanov, {\it Vanishing
 Theorems on Hermitian
Manifolds}, Diff. Geom. Appl. Vol {\bf 14} (2001) 251-265: math/9901090.}

\REF\chs{
C. G. Callan, Jr., J. A. Harvey and A. Strominger,
{\it Supersymmetric string solitons},
hep-th/9112030.}
 
 \REF\strom {A Strominger,
 {\it Superstrings with torsion}, Nucl. Phys.
{\bf B274} (1986) 253.}

\REF\candelas{P Candelas, GT Horowitz, A Strominger and E Witten,
{\it Vacuum configurations for superstrings}, Nucl. Phys. {\bf B258} (1985) 46.}

\REF\vc {
A.H.~Chamseddine and M.S.~Volkov,
{\it Non-Abelian BPS monopoles in N = 4 gauged
supergravity},
Phys.\ Rev.\ Lett.\  {\bf 79}, 3343 (1997)
:hep-th/9707176;
{\it Non-Abelian solitons in N = 4 gauged
supergravity and leading order  string theory},
Phys.\ Rev.\  {\bf D57}, 6242 (1998): hep-th/9711181.}

\REF\maldnun{Juan M. Maldacena and Carlos Nunez,
 {\it Towards the large N limit of pure N=1 superYang-Mills}, 
  Phys. Rev. Lett. {\bf 86}:588-591, (2001): hep-th/0008001.} 

\REF\papts{ G Papadopoulos  and AA 
Tseytlin, {\it Complex geometry of conifolds and five-brane wrapped on two
sphere}, 
Class.Quant.Grav. {\bf 18}:1333-1354 (2001): hep-th/0012034.}

%%%%%%%%%%%%%%%%%%%%%%%%%%%%%%%%
\Pubnum{ \vbox{ \hbox{} } }
\date{ January 2002}
\pubtype{}

\titlepage

\title {\bf KT and HKT Geometries in Strings and in Black Hole Moduli Spaces}

\author{George PAPADOPOULOS}
\address{Department of Mathematics,
\break King's College London,
\break
Strand,
\break
London WC2R 2LS, U.K.}
%\andauthor{}
%\address{}

\abstract{Some selected applications of
KT and HKT  geometries  in string theory,  supergravity,
  black
hole moduli spaces and  hermitian geometry are reviewed. 
 It is shown that the moduli spaces of a large class of five-dimensional
supersymmetric black holes are HKT spaces. In  hermitian geometry,  it is shown
 that a compact, conformally balanced, strong KT manifold whose associated KT 
connection has holonomy contained in $SU(n)$ is Calabi-Yau. The implication 
of this result in the context of some string compactifications is explained.}

\endpage
\pagenumber=2
%\sequentialequations

\chapter{Introduction}

In physics, geometries with torsion a three form have found  applications in
string theory, in supersymmetric quantum mechanics [\coles, \stelle] and in
the investigation of geometry of black hole moduli spaces 
[\stelle, \stroma, \gutpapa, \stromb, \gutpapb]. ( For other applications
see  [\hr, \howepapc].) The
typical geometric structure that appears is a triplet $(M,g,H)$,
where $M$ is a n-dimensional manifold with a metric $g$ which in
addition is equipped with a three form $H$. Such  manifolds apart
from the usual Levi-Civita connection $\nabla$ associated with
the metric $g$ also admit two more metric connections $\nabla^\pm$
which have torsion $\pm H$. We shall refer to $(M,g,H,
\nabla^\pm)$ as {\it T-manifolds}. The emphasis is on the connections
$\nabla^\pm$ because  many applications in physics involve the
reduction of the holonomy of these connections to an appropriate
subgroup of $SO(n)$. 

In string theory, $M$ is the spacetime, $g$ is the
Lorentzian spacetime metric and $H$ is the (closed) three-form
associated with the NS$\otimes$NS field strength. In
supersymmetric quantum mechanics, $M$ is the manifold that a
supersymmetric particle propagates, $g$ is a Riemannian metric,
and $H$ is a three-form which appears in some fermion couplings.
In the case of black holes, $M$ is the black hole moduli space, $g$ is the
moduli metric and $H$ is a three-form on the moduli space.

In mathematics, geometries with torsion a three-form arise in the context
of hermitian manifolds  which are {\it not} K\"ahler. A hermitian manifold
is a triplet $(M,g,J)$ of a manifold $M$ with Riemannian metric $g$ and a
complex structure $J$ such that $g(JX, JY)=g(X,Y)$ for any vector fields
$X$ and $Y$. For such manifolds,
 the complex structure
is not parallel with respect to the Levi-Civita connection. However
it has been known for sometime (see for example [\yano]) that there is
{\it unique} connection $\hat\nabla$
with torsion a three-form $H$ such that
$\hat\nabla g=\hat\nabla J=0$, ie the metric and complex
structure are parallel with respect to $\hat\nabla$. We shall refer 
to $(M,g,J,\hat\nabla)$ as a K\"ahler manifold with torsion or KT manifold
 for short [\stelle].
On every complex manifold there is always a KT-structure. This is
because given a complex structure it is always possible to find a metric
which satisfies the hermiticity property. Given a hermitian metric and
a complex structure, one can construct a
unique connection $\hat\nabla$. We shall refer to $\hat\nabla$ as KT-connection.
If the torsion $H$ is closed, $dH=0$, the KT-structure on $M$ is called strong
otherwise it is called weak.

A hyper-K\"ahler manifold with torsion or HKT manifold $(M,g,J_r, \hat\nabla)$ 
is a  manifold with
hypercomplex structure\foot{A manifold is hypercomplex if it
admits three complex structures $J_r$ that obey the algebra of
imaginary unit quaternions $J_1^2=J_2^2=-1$ and $J_3=J_1 J_2= -J_2
J_1$.} $\{J_r; r=1,2,3\}$, a tri-hermitian metric $g$, $g(J_rX,
J_rY)=g(X,Y)$, and a metric connection $\hat\nabla$ with torsion a
three-form $H$ such that all three complex structures are
parallel with respect to $\hat\nabla$, $\hat\nabla J_r=0$. Clearly
the HKT structure on a hypercomplex manifold is the analogue of KT
structure on a complex manifold. However it is not known whether
every hypercomplex manifold can always admit a HKT structure
unlike the case of a complex manifold which always admits a KT
structure. If the torsion $H$ is closed, $dH=0$, the
HKT-structure on $M$ is called strong otherwise it is called weak
in analogy with the KT case. The definition of the HKT structure was given
in [\howepap] and various properties have been investigated in [\howepap, \stelle, 
\poona, \poonb].
Many examples of HKT manifolds have been found which include group manifolds
[\tvp] and homogeneous spaces [\opap, \dofi]. For generalizations see 
[\howepapb, \opap, \ivanov, \ivanovd].

In this paper, we shall begin with a summary of the main
properties of complex and hypercomplex  geometry with emphasis
on the KT and HKT structures.
Then we shall describe how the T-geometries that appear in supersymmetric
quantum mechanics and in string theory are related to KT and HKT geometries.
We shall present two main results the following:

 \item{\bullet}The moduli space of supersymmetric five-dimensional
 black holes which preserve four supersymmetries is a weak HKT manifold.

 \item{\bullet}Compact, strong, conformally balanced, 
 KT-manifolds for which the holonomy
  of the KT-connection is contained in $SU(n)$, 
  ${\rm hol}(\hat\nabla)\subseteq SU(n)$,
   are necessarily Calabi-Yau.

The definition of a conformally balanced hermitian manifold 
will be given in the next section.
An application of the latter result  in string theory 
is that there are no supersymmetric
 warped compactifications of the common sector of type 
 II string theory with non-vanishing
NS$\otimes$NS three-form and 
${\rm hol}(\hat\nabla)\subseteq SU(n)$.

The material that I  present on the geometry of black hole
moduli spaces is a selection of the work done in collaboration
with Jan Gutowski in [\gutpapa, \gutpapb, \gutpapc]. Most of the material that
 I describe on KT-manifolds
with holonomy ${\rm hol}(\hat\nabla)\subseteq SU(n)$ has been done in
collaboration with Stefan Ivanov in [\ivapapa, \ivapapb].

This paper has been organised as follows: In section two, a
summary of the main properties of hermitian, KT and HKT manifolds
is presented. In section three, the relation between
supersymmetric mechanics and geometries with torsion is
explained. In section four, it is shown that the moduli spaces of
five-dimensional black holes which preserve four supersymmetries
are HKT manifolds. In section five, the relation between type II
supergravity and geometries with torsion is explained.  In
section six, it is shown that a class of compact,
conformally balanced, KT manifolds with ${\rm hol}(\hat\nabla)\subseteq SU(n)$ are
  in fact Calabi-Yau and an application to string
compactifications is presented. In addition, a non-compact example
of a KT manifold with ${\rm hol}(\hat\nabla)\subseteq SU(n)$ is given.

\chapter{Hermitian, KT and HKT manifolds}

Let $(M,g,J)$ be a hermitian manifold. Using the hermiticity 
condition of the metric $g$,
  $g_{ij}=g_{k\ell}J^k{}_i J^\ell{}_j$, we can define
 a K\"ahler two-form $\Omega$ on $M$ as $\Omega_{ij}=g_{ik}J^k{}_j$.
 For hermitian
 manifolds which are {\it not} K\"ahler, $\Omega$ 
 is {\it not} closed,  $d\Omega\not=0$. 
 Observe that $dvol(M)={1\over 2^n} \Omega^n$
 and so $M$ is oriented;
 (${\rm dim}M=2n$).

 There are several connections on $M$ which 
 preserve the hermitian
structure, ie
 they have the property that both the metric and 
 complex structure are parallel.
 Here we shall focus on two such connections. One 
 is the Chern connection defined
 as
 $$
\tilde \nabla_i Y^j=\nabla_iY^j+{1\over2} J^m{}_i d\Omega_{mkn} g^{nj} Y^k\ ,
 $$
 where $Y$ is a vector field, $\nabla$ is the Levi-Civita connection associated
 with the metric $g$ and $i,j,k=1,\dots, {\rm dim}M$.
The torsion of this connection is $$ C_{ijk}={1\over2}
\big(J^m{}_i d\Omega_{mjk}- J^m{}_j d\Omega_{mik}\big)\ . $$ (We
have lowered the upper index of the torsion using the metric $g$.)
Observe that $\tilde\nabla g=\tilde\nabla J=0$. The main property
of the Chern connection is that the curvature two-form is (1,1)
with respect to $J$ and therefore $\tilde\nabla$ is compatible
with the complex structure of the tangent bundle of $M$.

The KT-connection $\hat\nabla$, which appears in physics applications,
is the unique hermitian connection which has as torsion a three-form.
 The connection $\hat\nabla$ is
$$ 
\hat\nabla_i Y^j=\nabla_iY^j+{1\over2} g^{jm} H_{mik} Y^k\ ,
\eqn\ktcon 
$$ 
where  $Y$ is a vector field and the torsion $H$ is
$$
H_{ijk}=-3 J^m{}_{[i}d\Omega_{jk]m}\ . \eqn\kttor
$$
(Again we
have lowered the upper index of the torsion using the metric $g$.)
Observe that $\hat\nabla g=\hat\nabla J=0$. Therefore
$(M,g,J,\hat\nabla)$ is a KT-manifold. For generic hermitian
structures, $dH\not=0$, and so the associated KT structures are
weak.

In what follows, we shall use a relation between the
curvature\foot{Our conventions for the curvature $R$ of a
connection $\nabla$ are as follows: $[\nabla_i, \nabla_j] Y^k=
R_{ij}{}^k{}_\ell Y^\ell$  and $R_{ijk\ell}=-g_{km}
R_{ij}{}^m{}_\ell$.} $\tilde R$ of Chern connection $\tilde
\nabla$ and curvature $\hat R$ of KT connection $\hat \nabla$.
For this
 define
$$ 
u=-{1\over4} \tilde R_{ijkl} \Omega^{ij} \Omega^{kl} 
\eqn\defu
$$
 and 
 $$ 
 b=-{1\over2} \hat R_{ijkl}\Omega^{ij}\Omega^{kl}\ .
\eqn\defb 
$$
 It has been shown in [\ivapapa, \ivapapb] that
 $$ 
  2u=b+C_{ijk} C^{ijk}+{1\over4} (dH)_{ijkl}\Omega^{ij}\Omega^{kl}\ .
 \eqn\rub 
 $$
This formula is valid for any hermitian manifold.

The {\it Lee} form of a hermitian manifold $(M,g,J)$ is defined as follows:
$$
\theta=-J^m{}_i\nabla^k\Omega_{km} dx^i\ .
\eqn\lee
$$
We say that $(M,g,J)$ is {\it conformally balanced}, if the
 Lee form $\theta$ is {\it exact}.
In the case that the Lee form {\it vanishes} $\theta=0$, $(M,g,J)$ 
is called {\it balanced}. It can be shown that
if $(M,g,J)$ is conformally balanced with $\theta=df$, then
$(M,e^{{f\over 1-n}}g,J)$ is balanced ($n>1$).

For HKT manifolds $(M,g,J_r, \hat\nabla)$, the connection
$\hat\nabla$ is defined for each complex structure as in \ktcon.
In particular, this implies that
 the three KT torsions \kttor\
associated with the three complex structures $J_r$ are equal.

Some  HKT geometries arise from a HKT potential [\poona, \stroma].
Let $(M, J_r)$ be a hypercomplex manifold and a function $\mu$ on
$M$. Then a HKT structure can be defined on $(M,J_r)$ as
$$
\eqalign{ ds^2(M)&= \big(\partial_i\partial_j+ \sum_{r=1}^3
(J_r)^k{}_i(J_r)^\ell{}_j \partial_k\partial_\ell\big)\mu\, dx^i
dx^j\cr H&=d_1d_2d_3\mu} \eqn\pottor
$$
provided that $\mu$ can be
chosen such that the HKT metric above is well defined, where
$d_r=i(\partial_r-\bar\partial_r)$; $\partial_r$ is the
holomorphic exterior derivative with respect to $J_r$ complex
structure.

Some of the applications of KT and HKT geometries in physics are as follows:

\item{\bullet}Strong KT and HKT geometries have applications in type II
 string theory and in
 two-dimensional supersymmetric sigma models [\strom, \hr, \howepapc].

\item{\bullet}Weak KT and HKT geometries have  applications in
supersymmetric quantum mechanics [\coles, \stelle].

\item{\bullet}Strong and weak HKT geometries have applications in the moduli
spaces of gravitational solitons and  black holes [\stelle, \stroma, \gutpapa].

\chapter{Supersymmetric mechanics}

The supersymmetry algebra in one-dimension is spanned by the
generators $\{Q_I, T; I=1,\dots, {\cal N}\}$, where $Q_I$ are the
supersymmetry generators and $T$ is the translation generator, subject to the
anti-commutator relation 
$$ Q_I Q_J+Q_J Q_I=2 \delta_{IJ} T\ .
\eqn\susya 
$$ 
Supersymmeric mechanical systems are those that are
invariant under (infinitesimal) symmetries which realize the above
algebra. There are different realizations of the above
supersymmetry algebra and have been investigated in [\coles].

\section{{\cal N}=1 supersymmetry}

To find a system invariant under one supersymmetry, let $(M,g,H,
\nabla^\pm)$ be a T-manifold $M$ equipped with metric $g$ and a
three-form $H$. Next consider a map $X:~\bR\rightarrow M$. A
class of {\cal N}=1 supersymmetric mechanics models\foot{The torsion three-form $H$
in supersymmetric mechanics is usually denoted with $c$.} can be
described by the action [\coles]
$$ 
\eqalign{ I={1\over2}\int_{\bR}\,
dt\, \big( g_{ij} \partial_tX^i \partial_tX^j&+i g_{ij}
 \lambda^i\nabla^+_t \lambda^j
 \cr &-
{1\over24} (dH)_{ijkl}\lambda^i\lambda^j\lambda^k\lambda^l\big)\ ,}
\eqn\actone 
$$ 
where $\lambda$ is a (worldline)  fermion on $\bR$
which geometrically can be described as section of the bundle
$S\otimes X^*TM$; $S$ is the spinor bundle over $\bR$ and $X^*TM$
is the pull-back of the tangent bundle of $M$ with respect to the
map $X$. In addition, $\nabla^+_t$ is the pull-back of the
connection $\nabla^+$ on $\bR$ with respect to $X$. So
$$
\nabla_t^+ \lambda^i=\partial_t\lambda^i+(\Gamma^+)^i_{jk}
\partial_tX^j \lambda^k\ ,
$$
where $(\Gamma^+)^i_{jk}=\Gamma^i_{jk}+{1\over2} H^i{}_{jk}$ and $\Gamma^i_{jk}$
is the Levi-Civita connection of $g$.

The action \actone\ is supersymmetric because it can be written
in terms of superfields $X$ as
$$
I=-{1\over2}\int_{\Xi}\, dt\,
d\theta\, \big(i g_{ij} DX^i \partial_tX^j+{1\over12} H_{ijk} DX^i
DX^j DX^k\big)\ ,
 \eqn\acttwo
$$
where $D^2=i\partial_t$,
$D=\partial_\theta+i\theta\partial_t$, and $X: \Xi\rightarrow M$;
$\Xi$ is a supermanifold with an even coordinate $t$ and an odd
coordinate $\theta$. The infinitesimal supersymmetry
transformation is $\delta X^i=\eta QX^i$, where
$Q=\partial_\theta-i\theta\partial_t$ and $\eta$ is the
parameter; $DQ+QD=0$. The action \acttwo\ is invariant under this
supersymmetry transformation because it is a full superspace
integral.

To derive the action \actone\ from \acttwo, we 
integrate over the odd coordinate
$\theta$ which is equivalent to differentiating 
with respect to $D$ and evaluating the
resulting expression at $\theta=0$. The maps $X$ 
and fermions $\lambda$ in \actone\
are given in terms of the superfields\foot{It is customary to denote
the superfield and its first component with the same symbol.}
 $X$ as $X^i=X^i|_{\theta=0}$ and
$\lambda^i=DX^i|_{\theta=0}$.

The {\cal N}=1 supersymmetric mechanics system described by the
action \actone\ or equivalently by \acttwo\ is not the most general
one with {\cal N}=1 supersymmetry. More general models have been
constructed in [\coles]; see also [\stelle].

\section{{\cal N}=2B and {\cal N}=4B supersymmetry}

It is expected that the dynamics of black holes at small
velocities, ie in the moduli approximation, is described by a
action similar to  \actone\ which however is invariant under at
four supersymmetries instead of one. For this,  we 
investigate the conditions for \actone\ to be invariant
under one and three additional supersymmetries.  The 
infinitesimal transformations
of the additional supersymmetries are most easily written in terms of
${\cal N}=1$ superfields $X$ as 
$$ 
\delta X^i=\eta^r (J_r)^i{}_j
DX^j\ , 
$$ 
where $\eta^r$ are the anti-commuting infinitesimal
parameters and $J_r$ are endomorphisms of $TM$; $r=1$ or $r=1,2,3$.

Requiring that the above transformations satisfy the
supersymmetry algebra \susya\ and leave the action \acttwo\
invariant, one finds the following:

\item{\bullet} For models with two supersymmetries (${\cal N}=2B$)\foot{
The letter \lq B' has been added to denote a particular
realization of the supersymmetry algebra with two supercharges
according to the terminology used in [\stelle].}, 
$(M,g,H)$ is  a KT manifold $(M,g,J, \hat\nabla)$, where $J=J_1$
and $\hat\nabla=\nabla^+$

\item{\bullet}For models with four supersymmetries (${\cal N}=4B$),
$(M,g,H)$  is a HKT manifold  $(M,g,J_r, \hat\nabla)$,
where $J_r$ is the hypercomplex structure and $\hat\nabla=\nabla^+$.

In fact the above described geometric conditions for a model to
 admit two or four supersymmetries are sufficient but no necessary.
The derivation of the above conditions as well as  some more general results
can be found in [\coles,\stelle].

\chapter{Black hole moduli spaces}

Supersymmetric black holes are black hole solutions of
supergravity theories which in addition admit a number of
 Killing spinors. Killing spinors are solutions
of Killing spinor equations and an example of such 
equations will be described in detail in section five.
Supersymmetric black hole solutions in supergravity theories
apart from the spacetime metric also involve non-vanishing
Maxwell and scalar fields. The mass
 of the black holes
is related to their charges which is a consequence of a BPS type
of condition. Several supersymmetric black holes can be
superposed together to form a static configuration because there
is a balance of forces acting on them. Although the presence of Maxwell and
scalar fields in the solution are essential for the existence for such
superpositions, in what follows we shall 
focus on the spacetime metric of a supersymmetric
system with $N$  black holes. 
The spacetime metric of a typical solution
which describes $N$ supersymmetric black holes in superposition
of a supergravity theory can be expressed as
$$ 
ds^2=-A^2(\bfx,
\bfy_A)dt^2+ B^2(\bfx, \bfy_A) |d\bfx|^2\ , 
\eqn\mbm 
$$ 
where $(\bfx, t)$ are the spacetime coordinates, $|\cdot |$ is the
Euclidean inner product in $\bR^k$ and $\{\bfy_A\in \bR^k;
A=1,\cdots, N\}$ are the positions of the black holes. Note that
the components $A^2, B^2$ of the metric depend of the space
coordinates $\bfx$ and the positions $\bfy_A$ of the black holes.
As we shall see for the description of many aspects of the 
geometry of black hole moduli spaces,
the details of the supergravity action that \mbm\ is a 
solution are not essential.

The moduli space $\cM^k_N$  of a (supersymmetric)  N-black hole solution 
is the space of positions
of black holes. This can be identified with the space of  N-particles in $\bR^k$, ie
$$
\cM^k_N=\times^N \bR^k-\Delta\ ,
$$
where $\Delta=\{(\bfy_1,\dots,\bfy_N)\in \times^N \bR^k;\, 
\bfy_i=\bfy_j, i\not=j\}$ is the
diagonal. The dimension of the moduli space is $kN$.
If the  black holes have the same masses and carry the 
same charges, then the metric \mbm\
is invariant under the action of the permutation 
group $\Sigma_N$ acting
on the positions of the black holes. For such black 
holes, the moduli space
is
$$
\tilde \cM^k_N=\cM^k_N/\Sigma_N
$$
which is the configuration space of N-indistinguishable particles in $\bR^k$.

The geometry on the moduli space of black holes that preserve four
supersymmetries\foot{This means that these solutions admit four
non-vanishing Killing spinors, see section five.} is expected to be that 
of the \lq target' manifold
of supersymmetric mechanical systems which are invariant under
the same number of supersymmetries. This is because the
symmetries of a supergravity solution are expected to be
realized as symmetries of the associated effective theory.  Thus
for black hole systems which have as an effective theory the supersymmetric
mechanics models presented in section three, it is expected 
that their moduli space is a HKT manifold.

The cases of interest are those of black holes in four and five spacetime
dimensions. The computation of the metric on the moduli space is done as follows:

\item{\bullet}The positions of the black holes $\bfy_A$
 are allowed to depend on time $t$.

\item{\bullet}The metric and the other fields, like 
Maxwell fields, are perturbed
by first order terms in the black hole velocities.

\item{\bullet}These perturbations of the fields are 
determined by using the field equations.

\item{\bullet}The moduli metric is read by substituting 
the perturbed solution into
the appropriate supergravity action and by 
collecting the quadratic in the velocity terms.

The actual computation of the metric on the 
black hole moduli space is long and
 complicated. However Gutowski and I found that for most multi-black hole
 solutions, those that preserve at least four supersymmetries, the
 black hole moduli metric
can be determined from the components of the spacetime metric by
a simple relation that will be described below [\gutpapb]. We have shown
this by an explicit computation for the electrically charged
black holes of five- and four-dimensional supergravities which preserve
four superymmetries and are coupled to any
 number of Maxwell fields [\gutpapa, \gutpapb].
(Our results include the moduli metrics of the Reissner-Nordstr\"om  and the
graviphoton black holes which had been previously found in 
[\giba] and [\stroma], respectively.)
 We then conjectured that the same relation
between the spacetime metric and the moduli metric holds for all
systems of $N$-black holes
that preserve at least four supersymmetries. Our conjectured is
based on duality.

The metric on the moduli space of black holes can be determined from the
associated N-black hole spacetime metric as follows:

First define a function\foot{This function may not be well-defined
on the moduli space because the integral may not converge. However,
it can be shown that the moduli metric is  well-defined.}  
$\mu$,{\it ~ the moduli potential},
  on the moduli space $\cM^k_N$  ($k=3,4$) as
$$
\mu(\bfy_1, \cdots,\bfy_N)=\int_{\bR^k}d^kx\,\, 
A^{-2} B^2 (\bfx, \bfy_1, \cdots, \bfy_N)\ .
\eqn\hktpot
$$
The metric on the moduli space of four- and 
five-dimensional black holes can be determined
from $\mu$. In particular for five-dimensional black holes, the metric on
$\cM^4_N$ is
$$
ds^2(\cM^4_N)=\big[\partial_{mA}\partial_{nB}
+\sum_{r=1}^3 (I_r)^k{}_m  (I_r)^\ell{}_n
\partial_{kA}\partial_{\ell B}\big] \mu\,\, dy^{mA} dy^{nB}\ ,
\eqn\modmetr
$$
where  $\{I_r; r=1,2,3\}$ is a constant 
hypercomlex structure on $\bR^4$ associated
say with a basis of self-dual two forms on $\bR^4$.

The moduli space $\cM_N^4$ is a HKT manifold. 
To show this, one has to find a
hypercomplex structure on $\cM_N^4$ and identify 
the moduli potential $\mu$
given above with the HKT potential in section two.
 A hypercomplex structure on the
moduli space is
$$
({\bf I}_r)^{mA}{}_{nB}= \delta^A{}_B (I_r)^m{}_n\ .
\eqn\hypmod
$$
It is clear that this hypercomplex structure is 
induced from that on $\bR^4$.
Comparing \modmetr\ and \pottor\ using \hypmod, we can conclude
that the moduli metric  \modmetr\
is a HKT metric with potential $\mu$.
The torsion on the moduli space is then given as in \pottor.
Therefore we have shown the following:

\item{\bullet}The moduli space,  $\cM_N^4$, of  five-dimensional
 black holes which preserve
four supersymmetries
is a HKT manifold whose geometry is determined 
by the HKT potential given in \hktpot.

The metric on the moduli space of four-dimensional black holes
can be determined in a similar way. In this case the 
geometry on the moduli space and
the associated supersymmetric classical mechanics system  are 
more involved  and they will not be presented here. 
For more details see [\stromb, \gutpapb].

\section{STU black holes}
A large class of black hole solutions which preserve four supersymmetries are those 
of the STU supergravity in five-dimensions
with eight supersymmetries. The bosonic fields of this supergravity are
a graviton (a metric), three Maxwell fields and two scalars. In what follows,
the details
of the action of the STU supergravity theory are not important.
The spacetime metric of the multi-black hole solution is
$$
ds^2=-(f_1 f_2 f_3)^{-{2\over3}} dt^2+(f_1 f_2 f_3)^{1\over 3} |d\bfx|^2\ ,
$$
where
$$
 f_i=h_i+\sum_{A=1}^N {\lambda_{iA}\over |\bfx-\bfy_A|^2}
 $$
 for $i=1,2,3$ which are harmonic functions on $\bR^4$.
  The constants $\{h_i; i=1,2,3\}$ are related to the asymptotic
 values of the two scalars of the theory and the constants $\{\lambda_{iA};
 i=1,2,3 ; A=1, \dots, N\}$ are interpreted as the charge of the $A$-th black-hole
 with respect to the $i$-th Maxwell gauge potential; $\{\bfy_A; A=1,\dots,N\}$
 are the positions of the black holes.
 It is clear the solution is not invariant under the 
 action of the permutation group
 $\Sigma_N$ acting on the positions of black holes unless the 
 charges $\lambda_{iA}$ are equal.

 The moduli potential for the black holes of the  STU model
is
 $$
 \mu=\int_{\bR^4}\, d^4x \, f_1 f_2 f_3\ .
 \eqn\stump
 $$
For the masses of the black holes to be positive and for the
 black hole spacetime metric not to have naked singularities, we take
 $h_i,\lambda_{iA}>0$.
 The moduli potential \stump\ gives rise to the  moduli metric [\gutpapb]
 $$
 \eqalign{
 ds^2&= V_3 \sum_A[h_2 h_3 \lambda_{1A}+h_1 h_3\lambda_{2A} +h_1 h_2 \lambda_{3 A}]
 |d\bfy_A|^2
 \cr
+& V_3 \sum_{A\not= B} [h_2 \lambda_{1A} \lambda_{3B}+h_1 \lambda_{2A} \lambda_{3B}
 +h_3 \lambda_{1A} \lambda_{2B}] { |d\bfy_{AB}|^2\over |\bfy_{AB}|^2}
 \cr
 +&{ V_3\over2} \sum_{\{A\not=B\},C} \tau_{ABC}
 |d\bfy_{AB}|^2
\big[ {1\over |\bfy_{AC}|^2 |\bfy_{AB}|^2}+
  {1\over |\bfy_{BC}|^2 |\bfy_{AB}|^2}-{1\over |\bfy_{AC}|^2 |\bfy_{BC}|^2}
  \big]
  \cr
  -&2 \sum_{A\not=B\not=C} \int\, d^4x\, \tau_{ABC}
  {[ (dy_{A}^{[m} dy_{B}^{n]})^-\over |\bfx-\bfy_C|^2}
 \partial_m \big({1\over |\bfx-\bfy_A|^2}\big)
 \partial_n \big({1\over |\bfx-\bfy_B|^2}\big)\ ,}
 \eqn\modstu
 $$
 where ${\bfy}_{AB}=\bfy_A-\bfy_B$, $V_3$ is the volume of
  the unit three-sphere, $( dy_{A}^{[m} dy_{B}^{n]})^-$ is 
  the anti-self-dual part of
 $dy_{A}^{[m} dy_{B}^{n]}$
 and
 $$
 \tau_{ABC}=[\lambda_{1A}\lambda_{2B}\lambda_{3C}
 + \lambda_{1C}\lambda_{2A}\lambda_{3B}
 +\lambda_{1B}\lambda_{2C}\lambda_{3A}]\ .
 $$
 The moduli metric has a free term for N particles, and
  two-and three-body velocity dependent interactions.
 Observe that part of the moduli metric that contains three-body interactions
 is not given explicitly since the last term in \modstu\
 involves an integration over the spatial coordinates
 $\bfx$. This term has been investigated in [\gutpapc]
 and it was found that it can be determined by 
 the one-loop three-point amplitude
 of a $\phi^3$ theory.
  We remark that if the masses of the black holes
 $$
 m_A=h_2 h_3 \lambda_{1A}+h_1 h_3\lambda_{2A} +h_1 h_2 \lambda_{3 A}
 $$
 are not equal, then the centre of mass motion does not decouple
 from the dynamics of the system. For a discussion of the properties
 of the two-body system see [\gutpapc].

The two-black hole moduli space $\cM^4_2$ 
 is geodesically complete. In the case
that the centre of mass decouples, the 
relative moduli space is the connected
sum of two $\bR^4$, $\bR^4\#\bR^4$.
One asymptotic region is associated with 
black holes at small separations
while the other is associated with
 black holes at large separations.
Thus $\cM^4_2=\bR^4\times(\bR^4\#\bR^4)$, where $\bR^4$ is
associated with the free motion of the centre of mass. For $N>2$, it is not
known whether $\cM^4_N$ is geodesically complete.

It has been demonstrated in [\gutpapb] that the STU black holes at 
small separations exhibit a superconformal structure following a
similar work for the graviphoton black hole in [\stroma]. For 
applications of the superconformal symmetry in HKT geometry see [\swann].

\chapter{Type II supergravity}

In physics, there has been much activity in 
understanding the soliton-like solutions 
of supergravity theories because of their 
applications in the investigation
of non-perturbative properties of string theory. One property
of such solutions is that they are 
 {\it supersymmetric}. This means that 
 these supergravity solutions
  satisfy in addition  a set of at most first-order 
in the spacetime derivatives equations 
acting linearly on a spinor $\epsilon$ for
some $\epsilon\not=0$. These
 are  called {\it Killing spinor} equations. The {\it non-vanishing}
 solutions $\epsilon$ of Killing spinor equations are called 
 {\it Killing spinors}. 
In supergravity theories, the Killing spinor equations arise
as the vanishing conditions of the supersymmetry transformations of
the fermions of the supergravity 
theories and $\epsilon$ is the supersymmetry
infinitesimal parameter. 
Some of the Killing spinor equations 
are {\it parallel transport} equations
for the spinor $\epsilon$ with respect to a
 connection  of a spin bundle of spacetime. 
The
integrability conditions of the Killing spinor equations imply some of
the field equations of the supergravity theory. The existence
of Killing spinors
 is closely related to the stability of a
supergravity solution against small fluctuations. The {\it number of supersymmetries}
preserved by a solution is the {\it number of linearly independent} Killing spinors.

The simplest supergravity system for which some of the Killing spinor
equations have a direct geometric interpretation as parallel transport
equations is that of the  common sector or NS$\otimes$NS sector
of type II ten-dimensional supergravities [\schwarz, \howewest, \west]. 
We shall focus on the
type IIA theory. The discussion for the common sector of type IIB
supergravity is similar. It can be easily
extended to the case of heterotic string as well.

The bosonic fields of common sector of type IIA supergravity
 theory are the spacetime
metric $g$, a closed three-form field strength $H$ $(dH=0)$ and a scalar
field $\phi$ called dilaton. The spacetime $(M,g,H)$ is therefore
a T-manifold.
The field equations of the common sector of type IIA supergravity are
$$
\eqalign{
R_{MN}-{1\over4} H^R{}_{ML} H^L{}_{NR}+2\nabla_M\partial_N\phi&=0
\cr
\nabla_M\big(e^{-2\phi} H^M{}_{RL}\big)&=0\ .}
\eqn\feq
$$
In fact there is an additional field equation that of the dilaton $\phi$ but
it is implied from the above two equations. Let $\{\Gamma^M; M=0,\dots,9\}$
be a basis of the Clifford algebra Cliff($\bR^{1,9}$), ie 
$\Gamma^M\Gamma^N+\Gamma^N\Gamma^M
=2g^{MN}$.
Then the Killing spinor equations\foot{We have  used the notation
$\Gamma^{M_1\dots M_k}=\Gamma^{[M_1}\dots \Gamma^{M_k]}$.} are
$$
\eqalign{
\nabla^\pm \epsilon_\pm&=0
\cr
\big(\Gamma^M\partial_M\phi\mp {1\over12} H_{MNR} \Gamma^{MNR}\big)\epsilon_\pm&=0\ ,}
\eqn\kleq
$$
where
$$
\nabla^{\pm}_M Y^N=D_MY^N\pm {1\over2} H^N{}_{MR} Y^R
$$
and $\epsilon_\pm$ are sections of the spin bundles $S_\pm$, respectively\foot{
The spin group $Spin(1,9)$ has two inequivalent irreducible sixteen-dimensional
spinor representations and $S_\pm$ are the associated bundles.}.
It is clear that the first two of the Killing spinor equations are parallel
transport equations for the connections $\nabla^\pm$. Since these connections
of the spin bundles $S_\pm$ are induced from the tangent bundle of the spacetime,
the investigation of the Killing spinor equations is greatly simplified. This
is not the case in general, some of Killing spinor equations of supergravity theories
are parallel transport equations of a spin bundle but typically the associated
connections are not induced from the tangent bundle of
 spacetime\foot{The Killing spinor
equations of $D=11$ supergravity are parallel transport equations with
respect to a connection of the spin bundle which is not induced from the
tangent bundle of spacetime.}.

According to the definition given in the beginning
of this section, a solution of the field equations \feq\ is supersymmetric if it
satisfies the Killing spinor equations \kleq\ for some 
(non-vanishing) Killing spinors
$\epsilon_\pm$.

\section{The NS5-brane}

We shall illustrate the relation between the 
holonomy of the connections
$\nabla^\pm$ and the number of supersymmetries 
preserved with a simple example.
Consider the NS5-brane  solution [\chs]  of IIA supergravity
$$
\eqalign{
ds^2&=ds^2(\bR^{1,5})+f ds^2(\bR^4)
\cr
H&=\star df
\cr
e^{2\phi}&=f\ ,}
$$
where $\star$ is the Hodge star operation in $\bR^4$ and
 $f=1+{Q_5\over |\bfx|^2}$, $\bfx\in \bR^4$; $Q_5$ is the
 charge (per unit volume) of the NS5-brane.

Since the spacetime metric, three-form field strength
 and dilaton do not depend
of the coordinates of $\bR^{1,5}$, the non-trivial 
part of the solution is
$$
\eqalign{
ds^2&=f ds^2(\bR^4)
\cr
H&=\star df
\cr
e^{2\phi}&=f\ .}
\eqn\nsfive
$$
This metric describes the geometry of a smooth 
four-dimensional manifold $M$ equipped
with a closed three-form, ie $(M,g,H)$ is a
 T-manifold. As $|\bf{x}|\rightarrow \infty$,
the metric \nsfive\ becomes that of $\bR^4$ while as
$|\bf{x}|\rightarrow 0$ the metric becomes 
that of $\bR\times S^3$. 
In fact $M$ admits two constant hypercomplex structures 
 $\{J_r; r=1,2,3\}$ and
$\{I_r; r=1,2,3\}$. One is associated
with a basis of self-dual two forms on $\bR^4$ and the other
with a basis of anti-self-dual forms on $\bR^4$, respectively.
It turns out that $\nabla^+ J_r=0$ and 
$\nabla^- I_r=0$. This implies that
the holonomy of both $\nabla^\pm$ connections is
 contained in $Sp(1)$. It turns out that their holonomy is precisely $Sp(1)$ 
and there are sixteen parallel
spinors. The remaining two Killing spinor 
equations in \kleq\ are satisfied 
without additional
conditions. So one concludes the following:

\item{\bullet}The NS5-brane  
admits two HKT structures
and preserves sixteen supersymmetries.

\section{Supersymmetric compactifications on KT-manifolds}

String theory is formulated in ten-dimensions. 
To relate it to physics in
(d+1) dimensions ($d<9$), one uses 
solutions of the type
$$
\eqalign{
ds^2&=ds^2(\bR^{1,d})+ds^2(X)
\cr
H&=H(y)
\cr
\phi&=\phi(y)\ ,}
\eqn\comp
$$
where $H$ and $\phi$ are a three-form and a function on $X$,
respectively; $X$ is a compact manifold. At low energies the
(d+1)-dimensional theory emerges as the 
fluctuations of the \comp\ geometry
in ten dimensions.
Again the non-trivial part of  the
geometry \comp\ is described by the T-manifold $(X, g, H, \nabla^\pm)$.
Such compactifications have been considered in [\strom]; Such compactifications
for which the dilaton is {\it not} constant are also called {\it warped}
 compactifications.

A special class of solutions of this type 
are those for which $H=0$ and $\phi$
is a constant. In such a case, the 
Einstein equations imply that
the Ricci tensor vanishes. Moreover 
the Killing spinor equations imply
that the Killing spinors are parallel 
with respect to the Levi-Civita
connection. For both field equations 
and Killing spinor equations
to have solutions, $X$ must be the
 product of suitable irreducible Riemannian
manifolds with holonomy in $SU(n)$ ($n=2,3,4$), $G_2$, $Sp(2)$ or $Spin(7)$.

We shall focus on the investigation of 
compactifications for which $H\not=0$
and the holonomy of one of the connections $\nabla^\pm$, say $\nabla^+$,
is a subgroup of $SU(n)$ (${\rm hol}(\nabla^+)\subseteq SU(n)$). The investigation
of the geometry of $X$ that follows is due to [\strom] but we use the
terminology of [\ivapapa]  to describe it.
Since ${\rm hol}(\nabla^+)\subseteq SU(n)$, there is
a $\nabla^+$-parallel spinor $\eta$ such that
$$
J^i{}_j=-i\eta^\dagger \Gamma^i{}_j\eta
$$
is an almost complex structure. Since $\eta^\dagger\eta$ is constant, 
we have normalized $\eta$ such that $\eta^\dagger \eta=1$.
 In fact it can be shown that $J$ is an {\it integrable}
complex structure, {\it parallel} with respect to $\nabla^+$, $\nabla^+J=0$, and the
 metric $g$ is {\it hermitian}
with respect to $J$. Therefore,
 we conclude the following:

\item{\bullet} If ${\rm hol}(\nabla^+)\subseteq SU(n)$, then $(X,g,J,\hat\nabla)$ 
is a KT
manifold with $\hat\nabla=\nabla^+$.

Next we consider the second Killing spinor equation given by
$$
\big(\Gamma^i\partial_i\phi-{1\over12} \Gamma^{ijk} H_{ijk}\big)\eta=0
$$
on the parallel spinor $\eta$. This equation implies an additional
condition. In particular multiplying the above Killing spinor equation
 and its conjugate with $\Gamma^m$, and using the definition of $J$, we find
that
$$
2\partial_i\phi-{1\over2} J^m{}_iH_{mjk} \Omega^{jk}=0\ ,
$$
where $\Omega$ is the K\"ahler form of $X$. In [\ivapapa, \ivapapb], it was observed 
using $\hat\nabla J=0$ that the
one-form
$$
\theta=J^m{}_iH_{mjk} \Omega^{jk} dy^i
$$
is the {\it Lee form} \lee, $\theta$, 
of the Hermitian manifold $X$ as defined in  \lee.
Thus we have that
$$
\theta_i=2\partial_i\phi\ .
$$
 Since the Lee form
is {\it exact}, $\phi$ is a real function on $X$, $X$ is a {\it conformally balanced} 
hermitian manifold.
 Therefore we find that supersymmetric compactifications
of type II strings for which $H\not=0$ and ${\rm hol}(\nabla^+)\subseteq SU(n)$ 
are associated with  manifolds $X$ which have the
following properties:

\item{\bullet}  $(X,g,J,\hat\nabla)$ is a compact, conformally balanced, 
strong KT-manifold whose
KT connection $\hat\nabla$ has holonomy which is a subgroup $SU(n)$.

An important property of the above manifolds is that they admit
a non-vanishing holomorphic (n,0)-form [\strom]. To be precise, we have the
following: 

\item{\bullet} Let $(X,g,J,\hat\nabla)$ be a conformally balanced KT-manifold
 with ${\rm hol}(\hat\nabla)\subseteq SU(n)$,
then $X$ admits a holomorphic (n,0) form.

To prove this, since ${\rm hol}(\hat\nabla)\subseteq SU(n)$, 
there is a parallel (n,0)-form
$\alpha$, $\hat\nabla\alpha=0$. Since $X$ is conformally balanced, then
the Lee form can be written as $\theta=2d\phi$ for some function
$\phi$ on $X$.  Then the (n,0)-form 
$\tilde\alpha=e^{-2\phi}\alpha$ is holomorphic.
We can easily demonstrate this by 
computing $\bar\partial\tilde\alpha$ using
the fact that $\alpha$ is parallel and $\theta=2d\phi$.
 \rightline{$\diamond$}

This concludes the description of the geometry of the manifolds $X$ that
arise in the compactifications which have been investigated in this section.
In the next section we shall address the question whether such manifolds can exist.

\chapter{Weakly Balanced KT-manifolds with ${\rm hol}(\hat\nabla)\subseteq SU(n)$}

To find whether compact, conformally balanced, strong KT-manifold with
${\rm hol}(\hat\nabla)\subseteq SU(n)$ can exist, we  define the 
holomorphic Laplace operator on a function $f$ as
$$
L(f):=-2g^{\alpha\bar\beta} \partial_\alpha\partial_{\bar\beta}f
$$
and observe that it can be rewritten\foot{This corrects a
misprint in [\ivapapa].} as
$$
L(f)=\Delta f+g^{ij}\theta_i \partial_jf \ ,
$$
where $\Delta=-\nabla^i\partial_i$ is the standard Laplace operator.
The main result shown in [\ivapapa, \ivapapb] is the following:

\item{\bullet} Let $(X,g,J,\hat\nabla)$ be a compact, strong, 
conformally balanced, KT-manifold 
with ${\rm hol}(\hat\nabla)\subseteq SU(n)$,
then $X$ is a Calabi-Yau manifold.

To show this, assume that  $(X,g,J)$ in non-K\"ahler. {}From the assumptions
of the theorem and the results of the previous section,
 $X$ admits a holomorphic
(n,0)-form $\tilde \alpha$.  Set 
$f=-{1\over2} |\tilde\alpha|^2$, where $|\cdot|$
is the norm with respect to the metric $g$. Then
$$
L(f)=-{1\over2} \Delta |\tilde\alpha|^2-{1\over2}g^{ij}
  \theta_i \partial_j|\tilde\alpha|^2\ .
 $$
 On the other hand using the holomorphicity 
 of $\tilde \alpha$, we find that
 $$
 L(f)=2 u |\tilde \alpha|^2+|\tilde\nabla\tilde \alpha|^2\ ,
 $$
 where $u$ is defined in \defu, section two.
 Next observe that
 $$
2u=C_{ijk} C^{ijk}>0
\eqn\pos
$$ 
and so $u>0$ for a KT but non-K\"ahler manifold. 
This follows from the \rub\ and the assumptions
 of the theorem which imply that $b=0$ because 
 ${\rm hol}(\hat\nabla)\subseteq SU(n)$ 
 and $dH=0$ because $X$ is strong KT manifold. 
 
Since $u>0$ and $\tilde \alpha\not=0$, $L(f)>0$. 
From the Hopf maximum principle follows that
 either $\tilde\alpha=0$ or $C=0$. Since 
 $\tilde\alpha\not=0$, it follows
 that $C=0$, ie the torsion of the Chern
  connection vanishes and  $X$
 is K\"ahler, so $X$ is a Calabi-Yau manifold. 
 \rightline{$\diamond$}

One could reach the same conclusion from \pos\
 using a Kodaira vanishing theorem. The above result can also
 be derived  under somewhat weaker assumptions[\ivapapa, \ivapapb]. However
 the proof  is more involved. Observe that if the manifold $X$ is {\it weak} KT,
 then it is not necessarily the case that $u>0$ because $dH\not=0$ in  \rub. This
 implies that  compact, weak, 
conformally balanced, KT-manifold 
with ${\rm hol}(\hat\nabla)\subseteq SU(n)$ can exist. For some
 more vanishing theorems
on hermitian manifolds see [\ivanovb].

An application of the above result [\ivapapa] to type
 II string theory  is the following:

\item{\bullet}The only smooth supersymmetric 
compactifications of common sector of
 type II strings with $\rm{hol}(\nabla^+)\subseteq SU(n)$
  are the Calabi-Yau
 compactifications of [\candelas] for which  $H=0$ 
 and the dilaton $\phi$ is constant.

In other words that are no such warped compactifications
of the common sector of type II string theory.
 In the case of heterotic string though, it may
 be possible to find  compactifications
with $H\not=0$
because $dH\not=0$ due to the anomaly
 cancellation. Therefore the relevant
compact manifolds are weak KT and as we 
have mentioned such smooth manifolds can exist.

\section{An  example of non-compact, KT-manifold
 with ${\rm hol}(\hat\nabla)\subseteq SU(3)$}

The assumption that $X$ is compact in the 
 theorem of the previous section is necessary. This is because
there are {\it non-compact}, strong, conformally balanced, KT-manifolds
$(X,g,J,\hat\nabla)$ for which ${\rm hol}(\hat\nabla)\subseteq SU(n)$.
Such an example was found in [\vc] and interpreted in [\maldnun] as a
gravitational dual of pure $N=1$ supersymmetric Yang-Mills theory
in four dimensions. The geometric interpretation of the  IIA
supergravity solution below in terms of KT-geometry with
 ${\rm hol}(\hat\nabla)\subseteq SU(3)$ 
was given
in [\papts]. Let $d\sigma^i=-{1\over2}\epsilon^i{}_{jk} \sigma^j\wedge \sigma^k$
be a basis of left-invariant one-forms in $S^3$. The KT geometry is
$$
 \eqalign{ ds^2&=dr^2+e^{2g(r))}
\big(d\theta^2+\sin^2\theta d\varphi^2)+{1\over4} \sum_{i=1}^3
(\sigma^i-A^i)^2 \cr H&=-{1\over4}
(\sigma^1-A^1)\wedge(\sigma^2-A^2)\wedge (\sigma^2-A^2)+
{1\over4} \sum^3_{i=1} F^i\wedge (\sigma^i-A^i) \cr
e^{2\phi}&=e^{2\phi_o} {2 e^g\over \sinh r}\ ,} 
\eqn\cvs 
$$
where $(\theta, \varphi)$ are the usual angular coordinates on $S^2$, 
$r$ is a radial coordinate and $\phi_0$ is an integration constant. In addition
$$
A^1=a(r) d\theta\qquad A^2=a(r)\sin\theta d\varphi\qquad
A^3=\cos\theta d\varphi $$ 
and 
$$ \eqalign{ a&={2r\over\sinh r}
\cr e^{2g}&=r\coth 2r-{r^2\over\sinh^22r}-{1\over4}\ .}
$$

The manifold associated with \cvs\  is complete and admits a
strong, conformally balanced, KT-structure with ${\rm hol}(\hat\nabla)=
SU(3)$. The K\"ahler form [\papts] is
$$ \eqalign{
\Omega&={1\over2} dr\wedge (\sigma^3-A^3)+X(r)
e^g\big(\sin\theta(\sigma^1-A^1)\wedge d\varphi -(\sigma^2-A^2)\wedge
d\theta\big) \cr &+ P(r) \big(-{1\over4} (\sigma^1-A^1)\wedge
(\sigma^2-A^2)+e^{2g} \sin\theta d\theta\wedge d\varphi\big)\ ,}
$$
where
$$ 
P={\sinh4r-4r\over 2\sinh^22r}\qquad X=\sqrt{1-P^2}\ . 
$$
It would be of interest to construct other examples of
non-compact, conformally balanced, strong (smooth) KT manifolds with 
${\rm hol}(\hat\nabla)\subseteq SU(3)$ because of the applications
that they might have in supersymmetric gauge theories.
\leftline{~} 
\leftline{~} 
\leftline{~} 
\noindent{\bf Acknowledgments:}  It is a pleasure to thank Jan Gutowski and
Stefan Ivanov for many useful discussions and suggestions. I would
like to thank the organizers of Bonn workshop on \lq\lq Special
Geometric Structures in String Theory"  for their kind invitation
and their warm welcome at the conference. This work is partly
 supported by the PPARC grant PPA/G/S/1998/00613. I am supported by a University
Research Fellowship from the Royal Society.

\refout
\bye